\documentclass[epjST]{svjour}
\usepackage{graphics}
\newcommand{\dd}{\mathrm{d}}

\usepackage{color}
\newcommand{\blau}[1]{\textcolor{black}{#1}}

\begin{document}
\title{Holographic vector meson melting in a thermal gravity-dilaton background 
related to QCD}
\author{R. Z\"ollner\inst{1,2} 
\and B. K\"ampfer\inst{1,2}\fnmsep\thanks{\email{kaempfer@hzdr.de}}  }
\institute{Helmholtz-Zentrum  Dresden-Rossendorf, 01314 Dresden, Germany \and
Institut f\"ur Theoretische Physik, TU~Dresden, 01062 Dresden, Germany}
\abstract{
A holographic model of probe vector mesons (quarkonia) is presented, 
where the dynamical 
gravity-dilaton background is adjusted to the thermodynamics of 
2 +1 flavor QCD with physical quark masses.
The vector meson action is modified to account for various quark masses. 
We focus on the $\Phi$, $J/\psi$ and $\Upsilon$
meson melting in agreement with hadron
phenomenology in heavy-ion collisions at LHC, that is the formation of hadrons
at the observed freeze-out temperature of 155 MeV.
} 
\maketitle

\section{Introduction} \label{Intro}

Heavy-flavor degrees of freedom receive currently some interest as valuable probes of hot and dense strong-interaction matter produced in heavy-ion collisions at LHC energies. The information 
encoded, e.g.\ in quarkonia ($c \bar c$, $b \bar b$) observables,  supplements
penetrating electromagnetic probes and hard (jet) probes and the rich flow observables,
thus complementing each other in characterizing the dynamics of quarks and gluons
up to the final hadronic states 
(cf.\ contributions in \cite{Proceedings:2019drx} for the state of the art). 
Since heavy quarks emerge essentially in early, hard processes, they witness
the course of a heavy-ion collision -- either as individual entities or subjects  of
dissociating and regenerating bound states 
\cite{Prino:2016cni,Yao:2018sgn}. 
Accordingly, the heavy-quark
physics addresses such issues as charm ($c$, $\bar c$) and bottom ($b$, $\bar b$)
dynamics related to transport coefficients 
\cite{Rapp:2018qla,Xu:2018gux,Cao:2018ews,Brambilla:2019tpt} 
in the rapidly evolving and highly
anisotropic ambient quark-gluon medium
\cite{Chattopadhyay:2019jqj,Bazow:2013ifa}
as well as $c \bar c$ / $b \bar b$
states as open quantum systems 
\blau{\cite{Katz:2015qja,Blaizot:2018oev,Brambilla:2017zei}}. 
The rich body of experimental data from LHC,
and also from RHIC, enabled a tremendous refinement of our theoretical treatment
which grew up on initiating ideas like
``Mott mechanism and the hadronic to quark matter matter phase transition" \cite{Blaschke:1984yj},
``$J/\psi$ suppression by quark-gluon plasma formation" \cite{Matsui:1986dk} and
``Statistical hadronization of charm in heavy ion collisions at SPS, RHIC and LHC"
\cite{Andronic:2003zv}.
For a recent survey on the quarkonium physics we refer the interested reader to \cite{Rothkopf:2019ipj}.  

The yields of various hadron species, light nuclei and anti-nuclei 
-- even such ones which are only very loosely bound -- 
emerging from heavy-ion collisions at LHC energies 
are described by the thermo-statistical model \cite{Andronic:2017pug}
with high accuracy. These yields span an interval of nine
orders of magnitude. Refinements have been proposed to resolve 
the so called proton puzzle and several parameterizations, 
e.g.\ related to excluded volume effects, account for
specific inter-hadron forces in the spirit of the Ma-Dashen-Bernstein theorem. 
The final hadrons and nuclear clusters are described by two parameters: 
the freeze-out temperature $T_{fo} =155$~MeV and a freeze-out volume depending on the centrality of the collision. Due to the
near-perfect matter-antimatter symmetry at top LHC energies the 
baryo-chemical potential $\mu_B$ is exceedingly small, $\mu_B/ T_{fo} \ll 1$. 
It is argued in \cite{Andronic:2017pug} that the freeze-out of color-neutral
objects happens just at the demarcation region of hadron matter to quark-gluon plasma,
i.e. confinement vs. deconfinement strong-interaction matter. 
In fact, lattice QCD results \cite{Borsanyi:2013bia,Bazavov:2014pvz} 
report a pseudo-critical temperature of 
$T_c = 155 \pm 9$~MeV, now further constrained to $T_c = 156 \pm 1.5$~MeV
\cite{Bazavov:2018mes} -- a value agreeing with the disappearance of the chiral condensates, a maximum of some susceptibilities and also
roughly with the minimum of sound velocity. 
The key is the adjustment of physical quark masses and the use of 2+1 
flavors, in short QCD$_{2+1}$(phys). Details of the (may be accidental)
coincidence of deconfinement and chiral symmetry restoration are matter of debate 
\cite{Suganuma:2017syi}, 
as also the formation of color-neutral objects out of the cooling quark-gluon plasma at $T_c$, in particular also very weekly bound nuclei. 
Note that at $T_c$ no phase transition happens, rather the thermodynamics is characterized by a cross-over, e.g.\ signaled by a pronounced dip in the sound
velocity which in turn is related to an inflection point of the pressure as a function of the temperature.

Among the tools for describing hadrons as composite strong-interaction
systems is holography. Anchored in the famous AdS/CFT correspondence,
holographic bottom-up approaches have facilitated a successful description of
mass spectra, coupling/decay strengths etc.\ of various hadron species.
While the direct link to QCD by a holographic QCD-dual or rigorous top-down
formulations are still missing, one has to restrict the accessible observables 
to explore certain frameworks and scenarios. We consider here a scenario 
which merges (i) QCD$_{2+1}$(phys) thermodynamics described 
by a dynamical holographic gravity-dilaton
background, where the notion ''dilaton`` refers to a bulk scalar field as conformal symmetry breaker, and (ii) holographic probe vector mesons.
We present a scenario which embodies QCD thermodynamics
of QCD$_{2+1}$(phys) and the emergence of hadron states at $T_c$  at the same time.
One motivation of our work is the exploration of a (sufficiently simple and thus
transparent) holographic model which is in agreement with the above hadron 
phenomenology in heavy-ion collisions at LHC energies. Early holographic attempts
\cite{Colangelo:2012jy,Colangelo:2009ra} to hadrons at non-zero temperatures faced the problem of meson melting
at temperatures significantly below the deconfinement temperature $T_c$.
Several proposals have been made \cite{Zollner:2016cgc} to find
rescue avenues which accommodate hadrons at $T_c$.

Even though we focus on charmonium and bottomonium we include -- for the
sake of comparison -- the $\Phi$ meson as strangeonium in our considerations. 
In the
temperature range $T \approx{\cal O} (T_c)$, the impact of charm and bottom
degrees of freedom on the quark--gluon-hadron thermodynamics is minor \cite{Borsanyi:2016ksw}. 
Thus, we consider quarkonia as test particles. We follow 
\cite{Gubser:2008ny,Finazzo:2014cna,Finazzo:2013efa,Zollner:2018uep} 
and model  the holographic background 
by a gravity-dilaton set-up, i.e.\ without adding further fundamental degrees
of freedom (as done, e.g.\ in \cite{Bartz:2016ufc,Bartz:2014oba})
to the dilaton, which was originally related solely to gluon degrees
of freedom \cite{Gursoy:2010fj}. That is, the dilaton potential is adjusted to QCD$_{2+1}$(phys)
lattice data. Such an approach looks quite different in relation to holographic
investigations of meson melting without reference to QCD thermodynamics
\cite{Braga:2016wkm,Braga:2016oem,Fujita:2009ca,Grigoryan:2010pj,Braga:2017bml}.
We restrict ourselves to equilibrium and leave non-equilibrium effects
\cite{Bellantuono:2017msk}
for future work.
 
Our paper is organized as follows. In section 2, the dynamics of the probe vector
mesons is formulated and its coupling to the thermodynamics related background
is explained.  (The recollection of the Einstein-dilaton dynamics is relegated to
Appendix A.) Numerical solutions in the strangeness
sector ($\Phi$) and charm ($J/\psi$) and bottom ($\Upsilon$) sectors
of vector meson states and the melting systematic are dealt with in section 3.
We summarize in section 4.  

\section{Probe vector mesons}

The action of quarkonia as probe vector mesons in string frame is 
\begin{equation} \label{eq:1}
S_m^V = \frac{1}{k_V} \int \dd^4x \, \dd z \sqrt{g_5} e^{-\phi} \, G_m(\phi) \,F^2 ,
\end{equation}
where the function $G_m(\phi(z))$ carries the flavor (or quark-mass, 
symbolically $m$)
dependence
and $F^2$ is the field strength tensor squared of a U(1) gauge field ${\cal A}$
in 5D
asymptotic Anti-de Sitter (AdS) space time with the bulk coordinate $z$ 
and metric fundamental determinant $g_5$; 
$\phi$ is the scalar dilaton field (with zero mass dimension).  
The structure of (\ref{eq:1}) is that of a field-dependent gauge kinetic term,
familiar, e.g., from realizations of a localization mechanism in 
brane world scenarios
\cite{Chumbes:2011zt,Eto:2019weg,Arai:2017lfv}.
In holographic Einstein-Maxwell-dilaton models (cf.~\cite{DeWolfe:2010he}), 
often employed in including a conserved charge density (e.g.~\cite{Rougemont:2015wca,Knaute:2017opk}),
such a term refers to the gauge coupling. For the relation to truncated string (M) theory and
supergravity, see remarks in \cite{Gubser:2009qt}.

The action (\ref{eq:1}) with $G_m = 1$, originally put forward 
in the soft-wall (SW) model for light-quark mesons \cite{Karch:2006pv}, 
is also used for describing heavy-quark vector mesons 
\cite{Braga:2016wkm,Braga:2016oem,Fujita:2009ca},
e.g.\
charmonium \cite{Grigoryan:2010pj,Braga:2017bml} or bottomonium. As emphasized, e.g.\ in \cite{Grigoryan:2010pj}, 
the holographic background encoded in $g_5$ and $\phi$ must be chosen
differently to imprint the different mass scales, since (\ref{eq:1}) with $G_m =1$
as such would be flavor blind. 
Clearly, the combination $\exp\{- \phi\} G_m(\phi)$ in (\ref{eq:1})
with flavor dependent function $G_m(\phi)$
is nothing but introducing effectively a flavor dependent dilaton profile
$\phi_m = \phi - \log G_m$,
while keeping the thermodynamics-steered hadron-universal dilaton $\phi$.
In fact, many  authors use the form 
$S_m^V = \frac{1}{k_V} \int \dd^4x \, \dd z \sqrt{g_5} e^{-\phi_m} \,F^2$
to study the vector meson melting by employing different parameterizations
of $\phi_m$ to account for different flavor sectors. Here, we emphasize
the use of a unique gravity-dilaton background for all flavors 
and include the quark mass (or flavor) dependence solely in $G_m$.

Our procedure to determine $G_m$ is based on the import of information from
the hadron sector at $T = 0$.  
The action  (\ref{eq:1}) leads via the gauges ${\cal A}_z  = 0$ and 
$\partial^\mu {\cal A}_\mu = 0$ and the ansatz
${\cal A}_\mu = \epsilon_\mu \, \varphi (z) \, \exp\{ i p_\nu x^\nu \}$
with $\mu, \nu = 0, \cdots, 3$ and the constant polarization vector $\epsilon_\mu$
to the equation of motion 
\begin{equation}
\varphi'' +
\left[\frac12 A' + (\partial_\phi \log G_m -1 ) \phi' + (\log f)' \right] \varphi' +
\frac{p^\mu p_\mu}{f^2} \varphi = 0,
\end{equation}
where a prime denotes the derivative w.r.t.\ the bulk coordinate $z$, $A$ denotes the warp factor and $f$ is the blackness function of the metric, cf. (\ref{eq:3}) below.
Via the transformation 
$\psi (\xi) = \varphi(z (\xi)) \, \exp\{ \frac 12 \int_0^\xi dz  \, {\cal S}_T (\xi) \}$ \blau{and by introducinbg the tortoise coordinate $\xi$ via  $\partial _\xi = (1/f) \partial_z$}
one gets eventually the form
of a one-dimensional Schr\"odinger equation with 
the tortoise coordinate $\xi$
\begin{equation} \label{eq:6}
\left[\partial_\xi^2 - (U_T(z(\xi)) - m^2) \right] \psi (\xi)= 0, \quad i = 0, 1, 2,  \cdots
\end{equation} 
with the mass $m$ in the rest system of the particle given by $m^2 = p^\mu p_\mu$. Normalizable solutions in the sense of $\int \nolimits_{0}^{\infty} |\psi(\xi)|^2 \mathrm{d}\xi=1$ 
require the boundary conditions $\psi(0)=0$ and $\psi_i(z\to z_H)=\psi(\xi \to \infty)\to 0$. A detailed analysis of the near-horizon behavior of the solutions of (\ref{eq:6}) shows that only the 
trivial solution $\psi(\xi)=0$ fulfills the condition $\psi(z \to z_H=0)=0$ if $m^2$ is a real number and $f \neq 1$, i.e.~there are no normalizable solutions except in the case of vanishing 
temperature. 
The Schr\"odinger equivalent potential is  
\begin{equation} \label{eq:7}
U_T \equiv \left( \frac12 {\cal S}_T' + \frac14 {\cal S}_T^2 \right) f^2 
+ \frac12 {\cal S}_T f f' 
\end{equation}
as a function of $\xi(z)$ with
\begin{equation} \label{eq:G}
{\cal S}_T \equiv  \frac12 A' - \phi' + \partial_z \log G_m (\phi(z)).
\end{equation}
At temperatures $T(z_H) > 0$, $f(z,z_H) \le 1$, $A = A(z, z_H)$, $\phi = \phi(z, z_H)$, and the
related thermodynamics is given as usual by the Einstein-dilaton bottom-up
model (see Appendix A for a recollection). Specifically, a cut-off $\xi_0$ is needed to ensure 
  \begin{equation} \label{eq:6a}
   \int \limits_{0^+}^{\xi_0} \! d \xi \, |\psi(\xi)|^2 =1.
  \end{equation}
We employ here $z(\xi_0)=z_H(1-\tilde \epsilon)$ with $\tilde \epsilon=10^{-2}$. \\
At $T = 0$ (label ``0''), $f = 1$ and $\xi = z$ and $U_T = U_0$ with
\begin{eqnarray} 
U_0 (z) & \equiv & \frac12 {\cal S}_0' + \frac14 {\cal S}_0^2 \label{eq:8}\\
{\cal S}_0& \equiv &  \frac12 A_0' (z) - \phi_0' (z) + 
\partial_z \log G_m (\phi(z)) , \label{eq:9}
\end{eqnarray}
and (\ref{eq:6}) becomes 
\begin{equation} \label{eq:S}
\left[ \partial_z^2 + (U_0(z) - m_i^2) \right] \psi_i = 0
\end{equation}
again with the normalizing conditions $\psi_i(0)=0$ and $\psi_i(z \to \infty) \to 0$\blau{, but this time without the necessity of introducing a cut-off}. \\
That is, at $T = 0$ one has to deal with a suitable Schr\"odinger equivalent 
potential $U_0 (z)$ to generate the wanted spectrum $m_i$. 
In such a way, the needed hadron physics information at $T = 0$ is imported
and included in $U_0$. The next step is solving (\ref{eq:8}) with the initial condition ${\cal S}_0 (z \to 0) \to -\frac1z$ (comming from the asymptotic AdS geometry in combination with 
(\ref{eq:9})) by numerical means
in general, since it is a Riccati equation, to obtain ${\cal S}_0 (z)$
and, with (\ref{eq:9}), then $G_m(\phi)$ with $G_m (0) = 1$.
This needs $A_0'(z)$ and 
$\phi_0'(z)$, which follow from the thermodynamics sector in Appendix A via
$A_0 = A(z) = \lim A_{z_H \to \infty}(z, z_H)$ and 
$\phi_0 = \phi(z) = \lim_{z_H \to \infty} \phi(z,z_H)$.
One has to suppose that these limits are meaningful.
The limited information from lattice QCD thermodynamics 
(both in temperature and in accuracy) may pose here a
problem. Ignoring such a potential obstacle we use then 
$G_m(\phi) = G_m(z(\phi_0))$ as universal (i.e.\ temperature independent)
function to analyze (\ref{eq:6}) with (\ref{eq:7}, \ref{eq:G}) at non-zero temperature.

\section{Systematic of quarkonia melting: 
Exploring a two-parameter ansatz of ${\mathbf U_0}$}

Our setting does not explicitly refer to a certain quark mass $m$. Instead,
an ansatz $U_0(z, \vec p)$ with parameter n-tuple $\{\vec p\}$
is used such to catch a certain quarkonium mass spectrum.
Insofar, $m$ is to be considered as cumulative label highlighting the dependence
of $G_m$ on a parameter set which originally enters $U_0$.
   
A simple two-parameter ansatz for $U_0(z)$ is 
\cite{Grigoryan:2010pj}
\begin{equation} \label{eq:10}
L^2 U_0 (z) = \frac34 \left( \frac{L}{z} \right)^2 + a^2  \left( \frac{z}{L} \right)^2
+ 4 b  
\end{equation}  
which is known to deliver via (\ref{eq:S}) the normalizable functions $\psi_i$ 
with discrete eigenvalues 
\begin{equation} \label{eq:11}
L^2 m_i^2 = 4 (a + b + i \,a), \quad i= 0, 1, 2, \cdots . 
\end{equation}
Clearly, (\ref{eq:10}) is a slight modification of the SW model ansatz 
\cite{Karch:2006pv} with 
$\frac{3}{4 z^2}$ stemming from the warp factor $A(z)$ and $z^2$
emerging from a quadratic dilaton profile.  
Note the Regge type excitation spectrum $m_i^2 = m_0^2 +  i a /L^2$ with
intercept and slope to be steered by two independent parameters $a$ and $b$. 
The level
spacing (referring to masses squared!) is $L^2 (m_1^2 - m_0^2) = 4 a$,
meaning $m_1^2 > 2 m_0^2$ for $a > \frac14 L^2 m_0^2$. In other words,
requesting small level spacing w.r.t.\ the ground state mass   
$m_0^2 = 4 (a + b) / L^2$ implies selecting smaller values of $a$.

As we shall demonstrate below, the ansatz (\ref{eq:10}) has several drawbacks and,
therefore, is to be considered only as illustrative example.
For instance, the sequence of radial  $\Upsilon$ excitations in nature does not form a linear Regge trajectory \cite{Ebert:2011jc}.
This prevents a unique mapping of $m_{0, 1} \to (a, b)$.
While the radial excitations of $J/\psi$ follow quite accurately a linear Regge trajectory in nature \cite{Ebert:2011jc},
the request of accommodating further $J/\psi$ properties in $U_0$ calls also for
modifying  (\ref{eq:10}), cf.\ \cite{Grigoryan:2010pj}. Despite the mentioned deficits,
the appeal of (\ref{eq:10}, \ref{eq:11}) is nevertheless the simply invertible relation $m_i^2 (a, b)$
yielding $a(m_{0,1})$ and $b(m_{0,1})$. Since we are going to
study the systematic, we keep the primary parameters $a$ and $b$ in what follows. 

Instead of discussing results at isolated points in parameter space
referring to $\Phi$, $J/\psi$ and $\Upsilon$ ground states $m_0$ and first
excited states $m_1$, we consider the systematic over the $a$-$b$ plane. 
We define as dissociation temperatures $T_{dis}^{g.s., 1st} (z_H)$ 
such values, at which the ground state or first excited state cannot longer be
accommodated in the Schr\"odinger equivalent potential $U_T$. 
At $T > T_{dis}^{g.s.}$ and $T > T_{dis}^{1st}$, the ground state or the
first excited state are called ``molten" or ``dissociated" since the respective
normalizable solutions of (\ref{eq:6}) \blau{in the sense of (\ref{eq:6a})} do not exist.
One might call $T_{dis}$ also break-up temperature.
We emphasize however as an aside that, in some significant temperature range 
above $T_{dis}^{g.s.}$,
the spectral functions of $J/\psi$ and $\Upsilon$ display still clearly
observably peaks despite the failure of $U_T$ to accommodate normalizable states
via the Schr\"odinger equation (\ref{eq:6}). These peaks, albeit not longer
spikes, obey thermal shifts and thermal broadening with increasing
temperature until approaching a smooth continuum at a certain
temperature $T_{peak-dis} > T_{dis}^{g.s.}$. 
Such a behavior could be interpreted
as pre-formation of modes which turn upon cooling into the
corresponding quasi-particles at lower temperatures.
We leave the investigation of spectral functions to separate follow-up
work and focus on implications of (\ref{eq:6}) - (\ref{eq:10}).

\begin{figure}[tb]
\center
\resizebox{0.99\columnwidth}{!}{%
\includegraphics{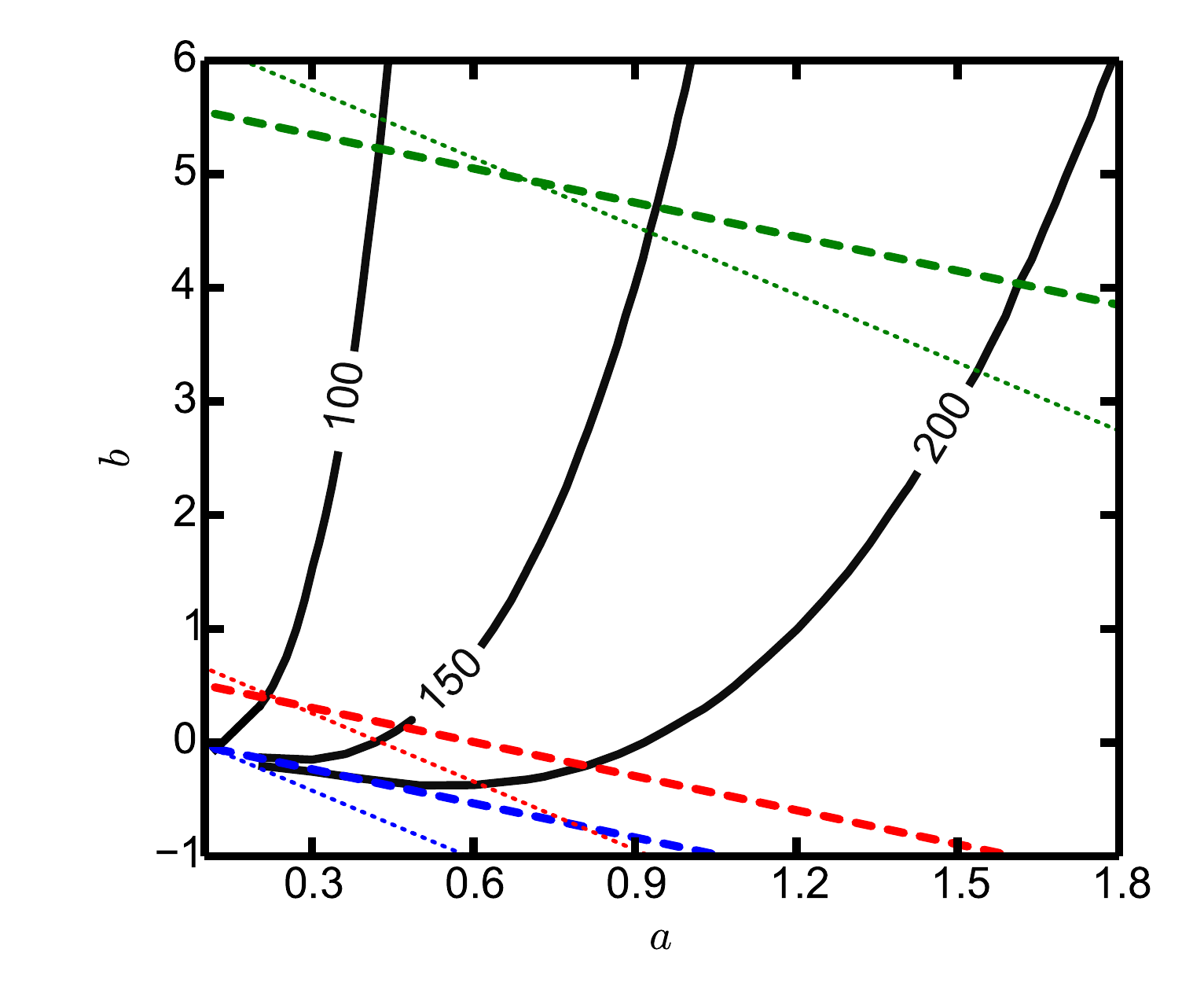} 
\includegraphics{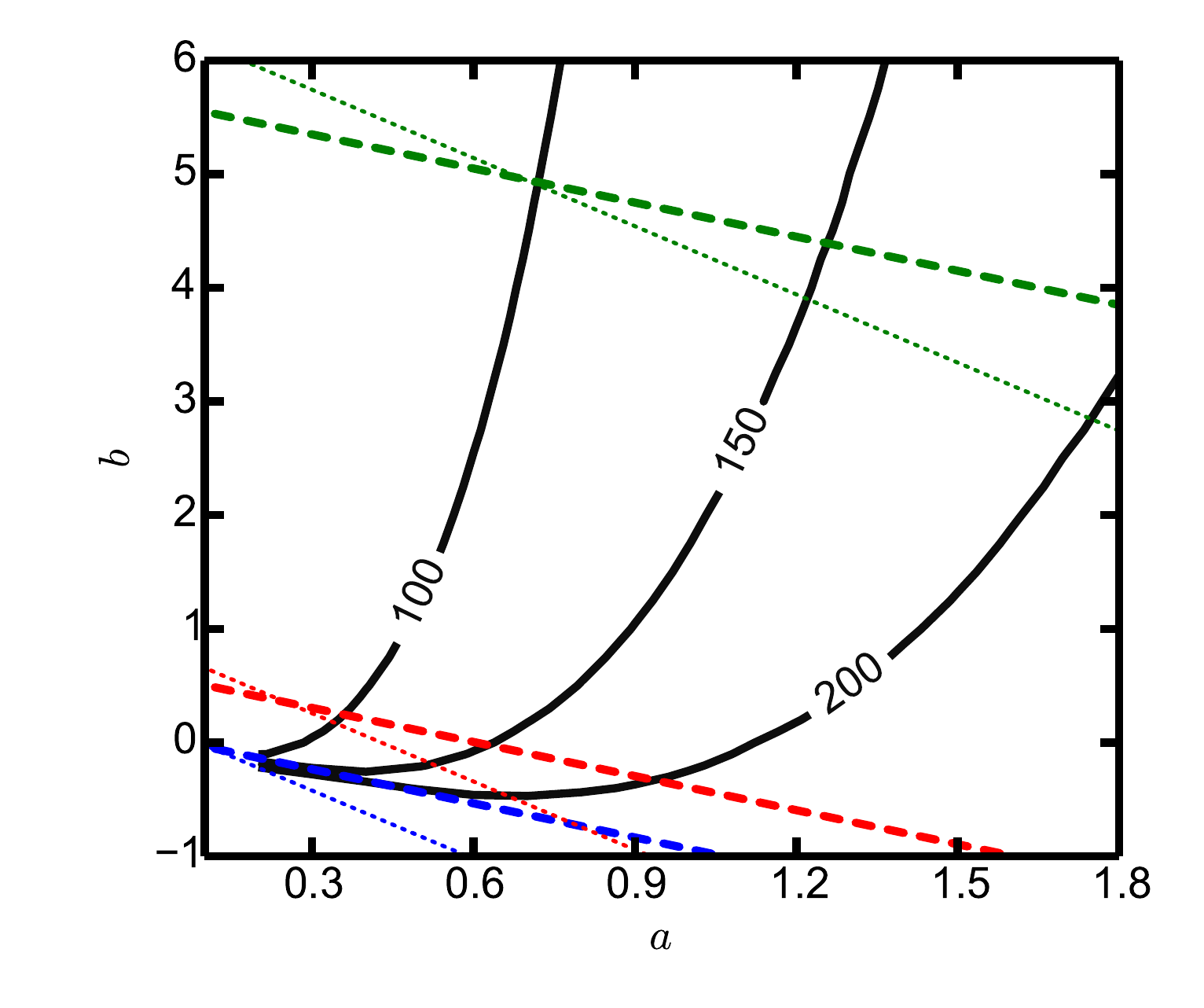} }
\caption{
Contour plot of $T_{dis}^{g.s.}$ (in MeV) over the $a$-$b$ plane
(solid labeled curves).
Also depicted are the curves $m_0 (a,b) = m_{g.s.} (\Phi, J/\psi, \Upsilon)$
(dashed) and $m_1 (a,b) = m_{1st} (\Phi, J/\psi, \Upsilon)$ (dotted)
with the color code $\Phi$-blue, $J/\psi$-red and $\Upsilon$-green.
Left panel: using the background adapted to QCD$_{2+1}$(phys) (see Appendix A),
right panel: using a schematic background with 
$A(z, z_H) = - 2 \log (z/L)$, $f(z,z_H) = 1 - (z/z_H)^4$, and
$\phi(z, z_H) =  (z/L)^2$.
Scale setting by $L^{-1} = 1.99$~GeV.
\label{fig:1}
}
\end{figure}

A survey of $T_{dis}^{g.s.} (a, b)$ is exhibited in Fig.~\ref{fig:1}-left by the
solid curves. The dashed curves depict the loci of $m_0 (a,b) = 1.019$~GeV ($\Phi$),
3.097~GeV ($J/\psi$) and 9.46~GeV ($\Upsilon$).
The two-parameter ansatz (\ref{eq:10})
allows for an independent fixing of $m_1 (a,b) = 1.68$~GeV ($\Phi_{1680})$,
3.686~GeV ($\psi(2S)$) and 10.0234~GeV ($\Upsilon (2S)$),
which adds a further congruence of curves (thin dotted). At the intersections
of the respective curves $m_0 (a,b)$ and $m_1 (a,b)$, the parameter doublets
$(a, b)$ would exactly correspond to the PDG values. It turns out, however,
that the dissociation temperatures would be too low in such a case. Instead of keeping
$m_1$ at PDG values, we allow for larger spacings and thus get larger
dissociation temperatures at given $m_0$.
Note that, according to the thermo-statistical
hadronization model \cite{Andronic:2017pug}, $T_{dis} \ge 155$ MeV
is required for all (ground state) hadron
species -- otherwise the impressive coverage of hadron abundances ranging
over nine orders  of magnitude would be hardly understandable.

Employing the QCD-related thermal background is quantitatively important.
If we use the schematic background parameterized by
$A(z, z_H) = - 2 \log (z/L)$, $f(z,z_H) = 1 - (z/z_H)^4$,
$\phi(z, z_H) =  (z/L)^2$, the quarkonia break-up curves
$T_{dis}^{g.s.}$ are shifted to right-down, see Fig.~\ref{fig:1}-right.
Nevertheless, the overall pattern is governed by the utilized ansatz $U_0(z; a, b)$.

\begin{figure}[tb]
\center
\resizebox{0.99\columnwidth}{!}{%
\includegraphics{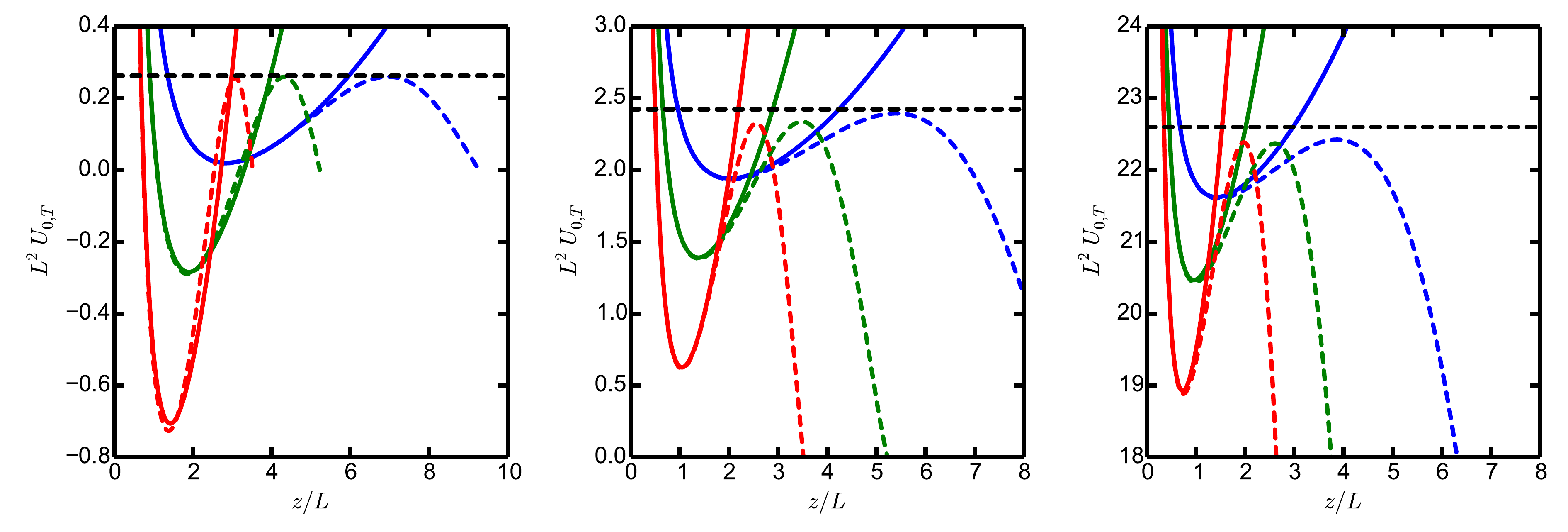} }
\caption{The Sch\"odinger equivalent potentials $U_0$ (solid curves) 
and $U_T$ (dashed curves) as a function of $z$
for $\Phi$ (left), $J/\psi$ (middle) and $\Upsilon$ (right).
The horizontal lines mark the respective ground states at $T = 0$. 
The parameters $(a, b)$ correspond to the crossing points of the curves
 $m_0 (a, b)$ with the dissociation temperatures $T_{dis}^{g.s.} = 100$~MeV
(blue), 150~MeV (green) and 200~MeV (red), cf.\ Fig.~\ref{fig:1}.
These temperatures are related via the relation $LT (z_H)$, 
exhibited in Fig.~\ref{fig:3}-right,
to the horizon position $z_H$. Note that $U_T(z, z_H)\vert_{z = z_H} = 0$
due to $f(z, z_H)\vert_{z = z_H} = 0$ in (\ref{eq:7}).
\label{fig:2}
}
 \end{figure}

To understand the reason of the variation of $T_{dis}^{g.s.} (a, b)$ let us consider
the Schr\"o\-dinger equivalent potentials $U_{0, T}$ for three points  $(a, b)$
in the parameter space which correspond to $T_{dis}^{g.s.} = 100$, 
150 and 200 MeV and given values $m_0 (\Phi, J/\psi, \Upsilon) = const$.    
The essence is that going on a curve $m_0 (a, b) = const$ to the right, i.e.\
keeping the ground state mass constant and enlarging $a$,
thus consecutively crossing the curves $T_{dis}^{g.s.} = 100$, 150, 200 MeV,
the Schr\"odinger equivalent potential $U_0$ and thus $U_T$ become
deformed in a characteristic manner from a shallow shape (small $a$) to a 
squeezed shape (large $a$), see Fig.~\ref{fig:2}. 
Thus, at a certain value of $z_H$ -- corresponding
to the temperature $T(z_H)$ -- the shallower Schr\"odinger equivalent potential
might not accommodate the ground state, while the squeezed and deeper
potential can do so. Analog features are met for excited states.
These observations offer an avenue for the design of suitably improved parameterizations
of $U_0$,
e.g.\ by introducing a pronounced dip at small values of $z$, as used in
\cite{Braga:2016wkm,Braga:2016oem,Fujita:2009ca,Grigoryan:2010pj,Braga:2017bml}.

\begin{figure}[tb]
\center
\resizebox{0.99\columnwidth}{!}{%
\includegraphics{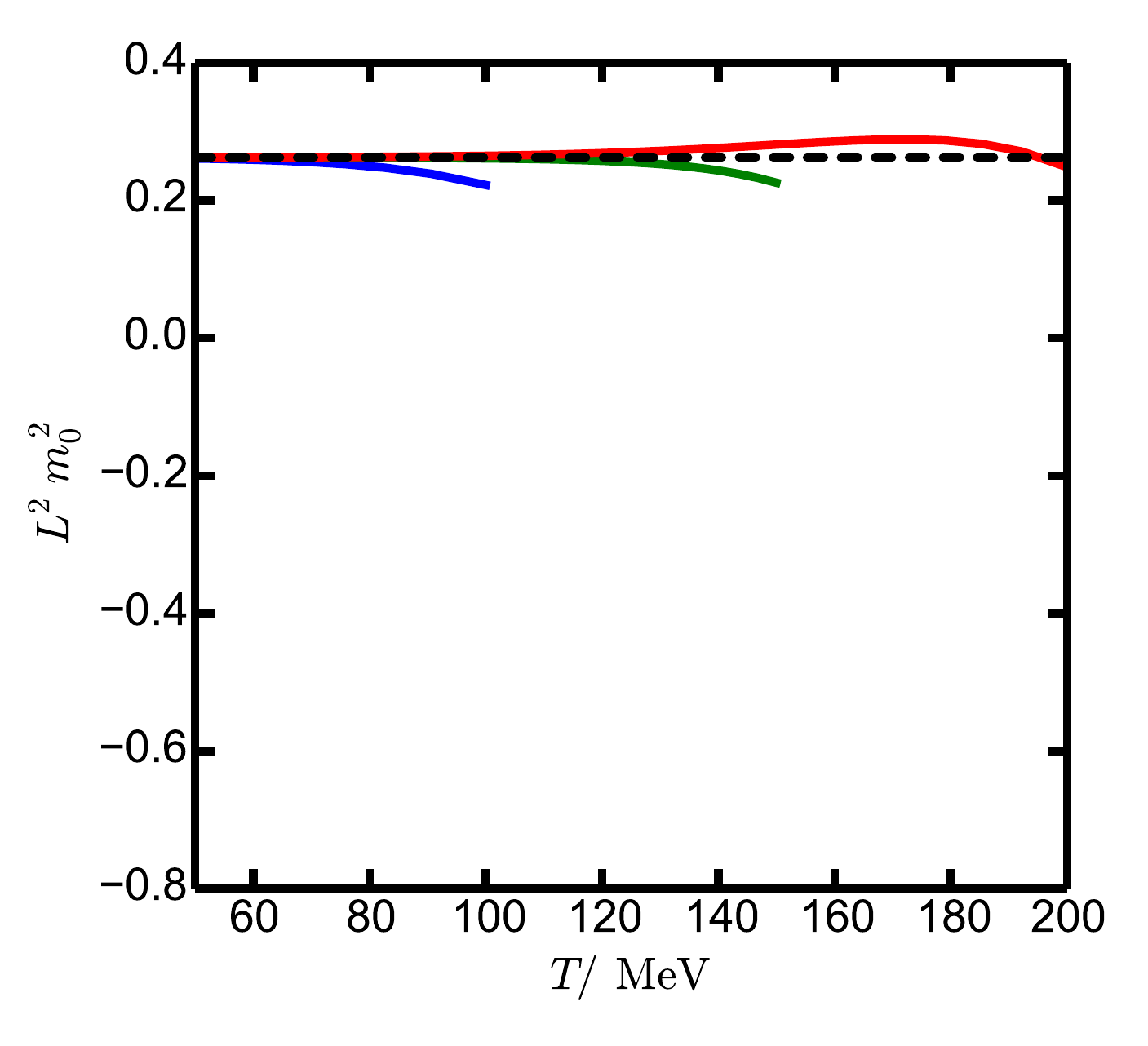}
\includegraphics{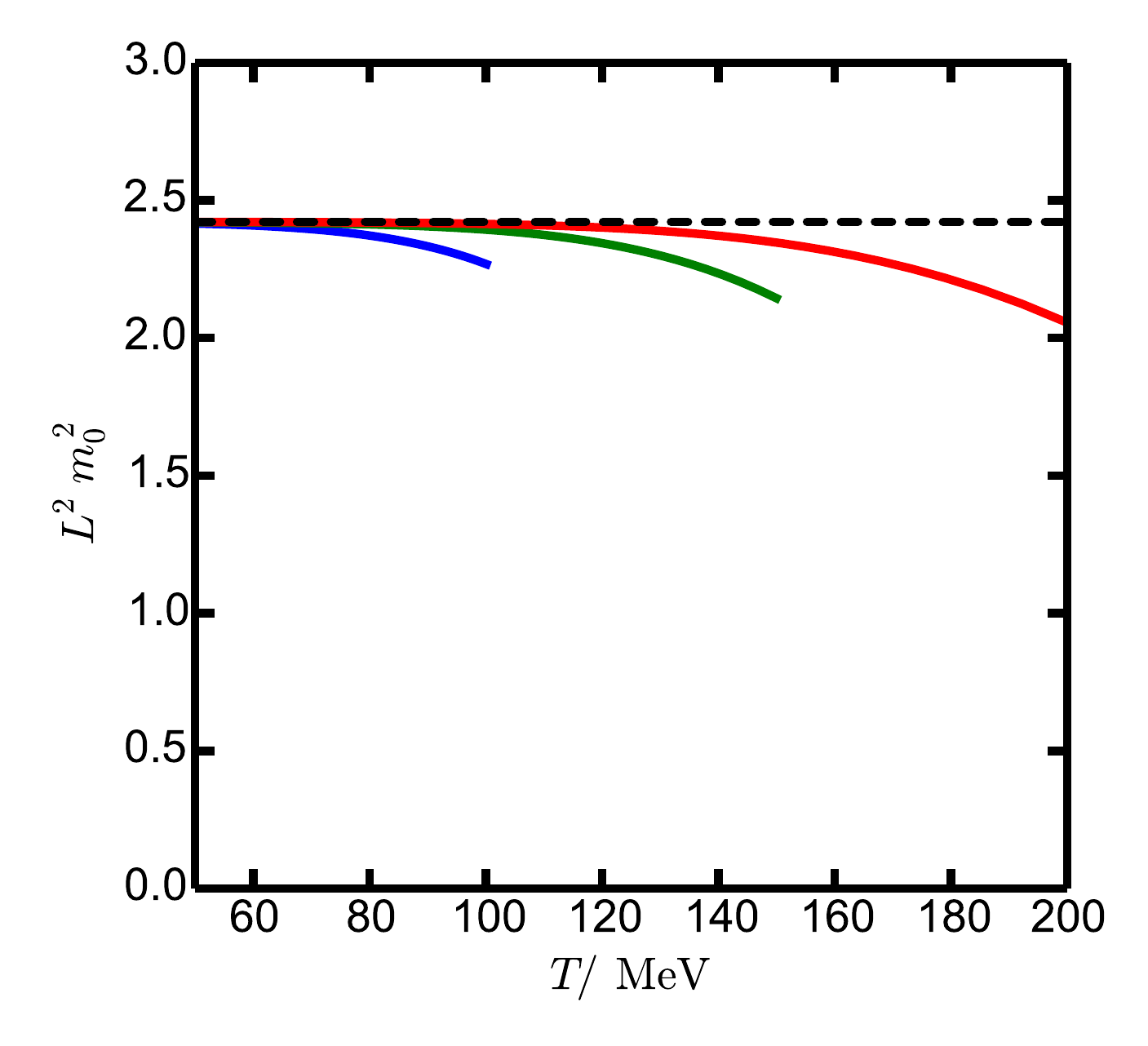} 
\includegraphics{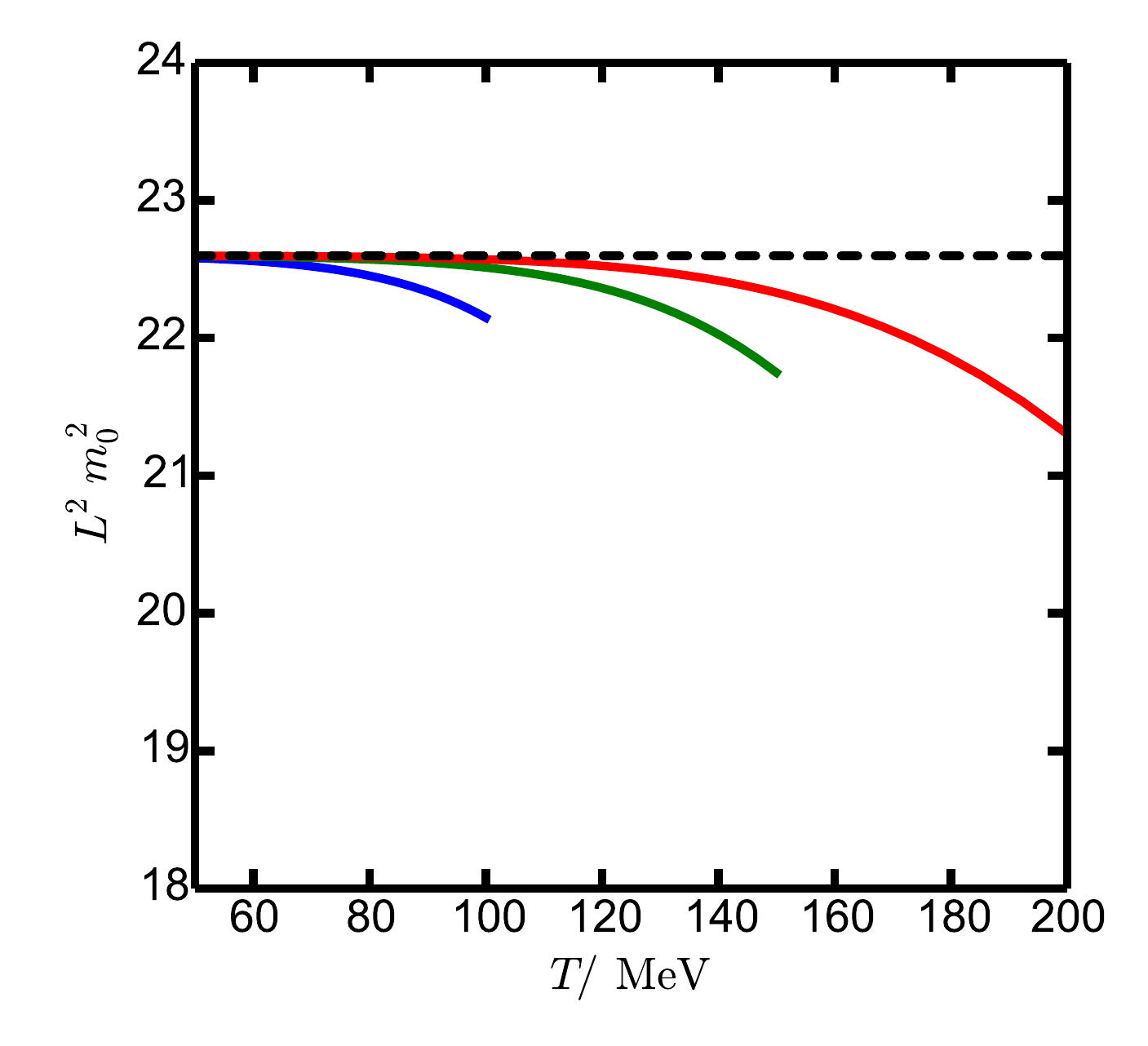} }
\caption{The dependence of the ground state masses $L^2m_0^2$
of $\Phi$ (left panel), $J/\psi$ (middle panel) and $\Upsilon$ (right panel) 
on the temperature $T$
for the same parameter doublets $(a, b)$ and color codes as Fig.\ \ref{fig:2}. 
The horizontal dashed lines depict the respective $T = 0$ values.
The same ordinate scales as in Fig.\ \ref{fig:2} are displayed.
\label{fig:7}
}
 \end{figure}

The ansatz
(\ref{eq:10}) facilitates a sequential melting upon increasing temperature,
$T_{dis}^{g.s.} > T_{dis}^{1st} > T_{dis}^{2nd}$ etc.,
and allows potentially for a strong flavor dependence 
$T_{dis}^{g.s.} (\Upsilon) > T_{dis}^{g.s.} (J/\psi) > T_{dis}^{g.s.} (\Phi)$.
Furthermore, the variation of $U_T$ as a function of the temperature
causes a {\sl negative} mass shift, see Fig.~\ref{fig:7}. Such thermal mass shifts
are employed in \cite{Brambilla:2019tpt} to pin down 
the heavy-quark (HQ) transport coefficient
$\gamma$ which can be considered as the dispersive counterpart
of the HQ momentum diffusion coefficient 
$\kappa = 2 T^3 /(D T)$, where $D$ stands for the HQ spatial diffusion
coefficient. Reference \cite{Rothkopf:2019ipj} stresses the tension within
previous holographic results \cite{Braga:2017bml}, where {\sl positive} mass shifts
are reported, in contrast to {\sl negative} shifts, e.g.\ in  \cite{Fujita:2009ca}.
Our set-up does not resolve that issue on a firm basis, since the thermal mass shifts 
of $J/\psi$ and $\Upsilon$ in 
Fig.~\ref{fig:7} are noticeably larger than the lattice QCD-based values 
quoted in \cite{Brambilla:2019tpt}. In addition, the sign of the thermal mass
shift can depend on the actual parameters $a$ and $b$ in the model 
Eq.~(\ref{eq:10}) as evidenced in the left panel of Fig.\ \ref{fig:7}. However,
the $\Phi$ meson should be considered neither as a HQ representative
nor as a proper probe due to missing back-reaction in our setting.

\begin{figure}[tb]
\center
\resizebox{0.99\columnwidth}{!}{%
\includegraphics{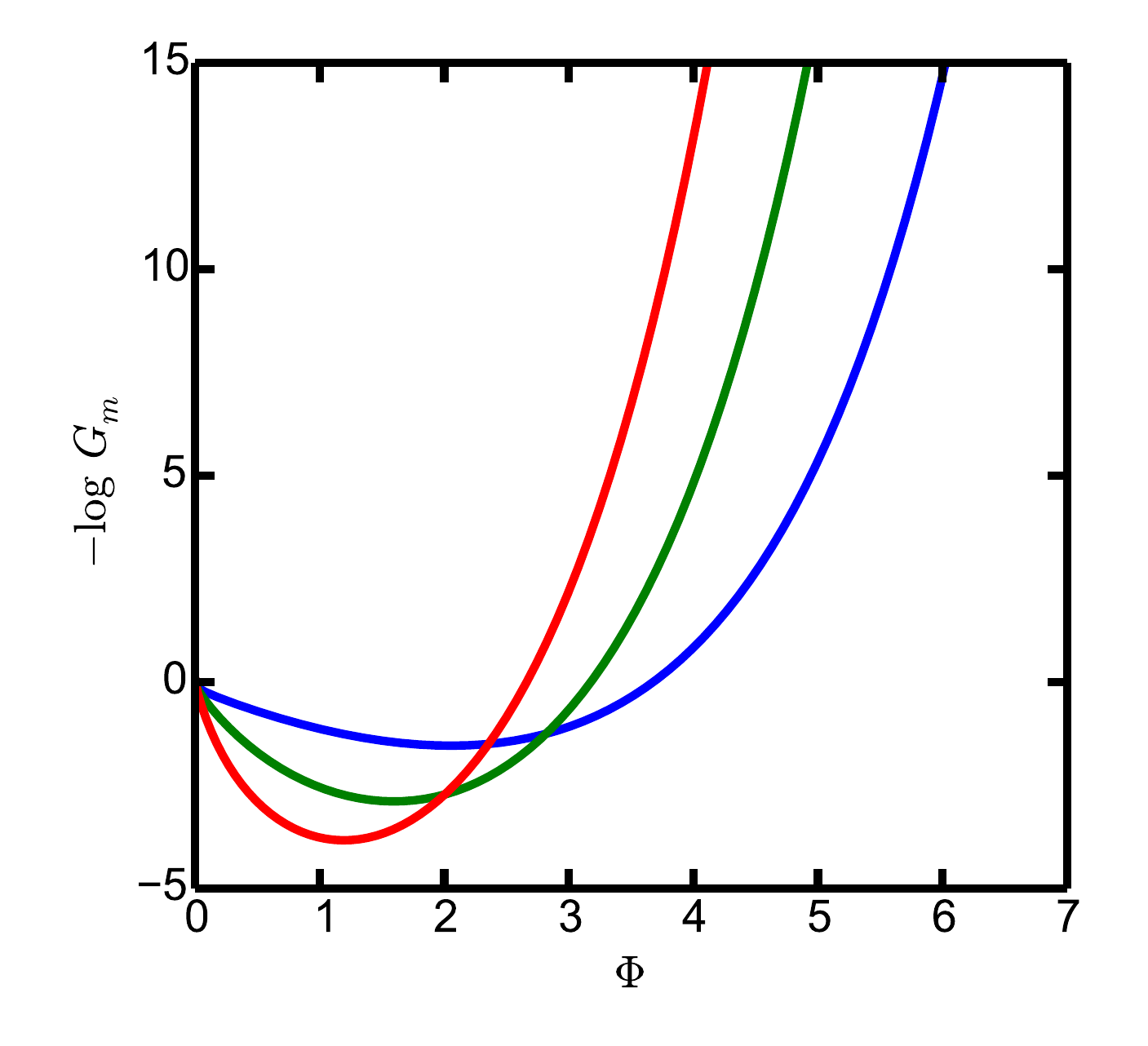}
\includegraphics{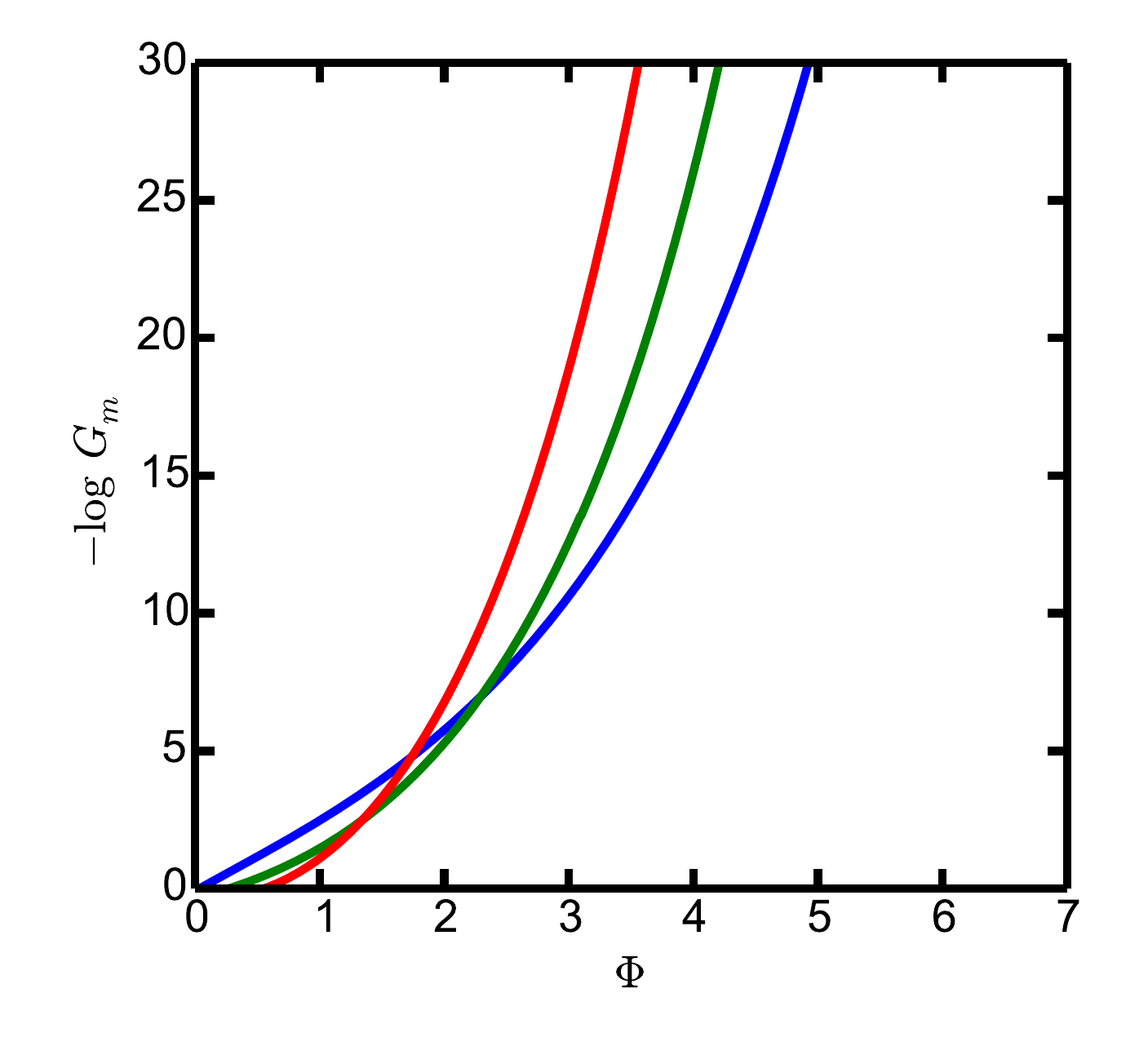} 
\includegraphics{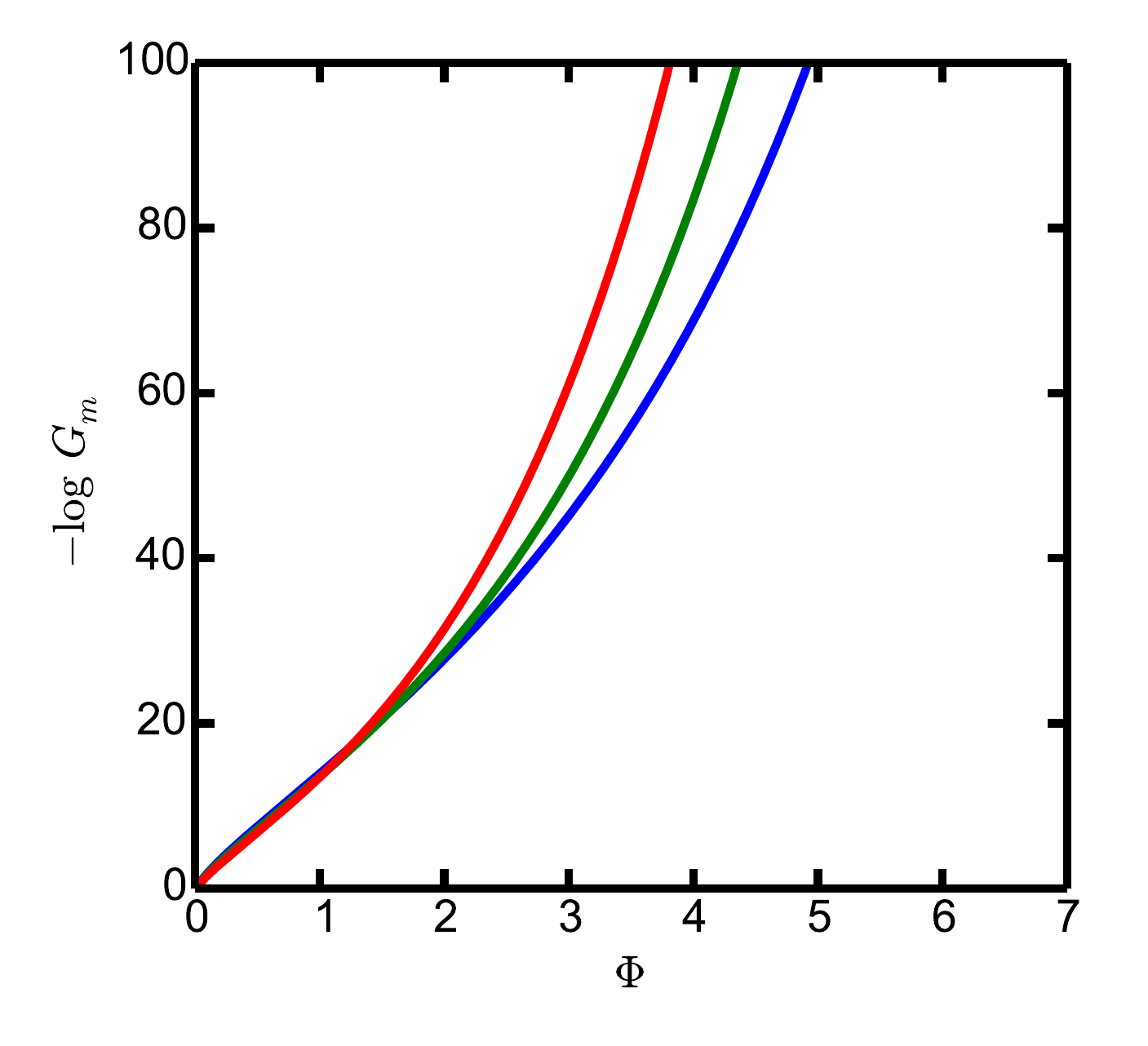} }
\caption{The quantities $- \log G_m (\phi)$ for the same parameter selection scheme
as in  Figs.\ \ref{fig:2} and \ref{fig:7}.
$\Phi$: left pane, $J/\psi$: middle panel, and $\Upsilon$: right panel.
\label{fig:6}
}
 \end{figure}

Finally, we exhibit in Fig.\ \ref{fig:6} the quantity
$- \log G_m$ as a function of $\phi$ with the same selection scheme of the
parameters $(a, b)$ as in Figs.\ \ref{fig:2} and \ref{fig:7}. 
Note the huge variation of $G_m (\phi)$ and the very strong flavor dependence. 
 
\section{Summary}

In summary we introduce a modification of the holographic vector meson action
for quarkonia
such to join (i) the QCD$_{2+1}$(phys) thermodynamics, described solely
by a dilaton and the metric coefficients, with (ii) a Regge type spectrum
at zero temperature. The vector mesons belong to different Regge
trajectories, e.g.\ the quarkonia $J/\psi$ and $\Upsilon$
and $\Phi$ as well.
The formal construction is
based on an effective dilaton $\phi_m = \phi - \log G_m$, where $\phi$ is
solely tight to the thermodynamics background, while the flavor dependent 
quantity $G_m$ is determined by
a combination of $\phi$ and the adopted Schr\"odinger equivalent potential.
The later one can be chosen with much sophistication to accommodate
many hadron properties. We use here only a two-parameter shape to
demonstrate features of our scheme, where the thermodynamic background 
at $T > 0$ and meson spectra at $T = 0$
serve as input to analyze the quarkonia melting at $T > 0$. According to
the hadron and nuclear phenomenology at LHC \cite{Andronic:2017pug}, 
hadrons must exist at and below temperatures $T_c \approx T_{fo}  \approx 155$ MeV.
Ideally, the light hadrons, represented in our approach by $\Phi$ mesons as
vector probe states, should form upon cooling at $T_c$, while charmonium 
or bottomonium must  have higher melting temperatures according to
QCD results \cite{Larsen:2019zqv,Kim:2018yhk}. 
Our holographic set-up is a purely static one, i.e.\
``meson melting'' is meant as a determination of the dissociation or break-up
temperature, rather than a dynamical process.

\begin{acknowledgement}

The authors gratefully acknowledge the collaboration with J.~Knaute
and thank
M.~Ammon, D.~Blaschke,
M.~Kaminski and K.~Redlich for useful discussions.
The work is supported in part by STRONG2020.

\end{acknowledgement}

\appendix
\section{Gravity-dilaton background }

\begin{figure}[b]
\center
\resizebox{0.99\columnwidth}{!}{%
\includegraphics{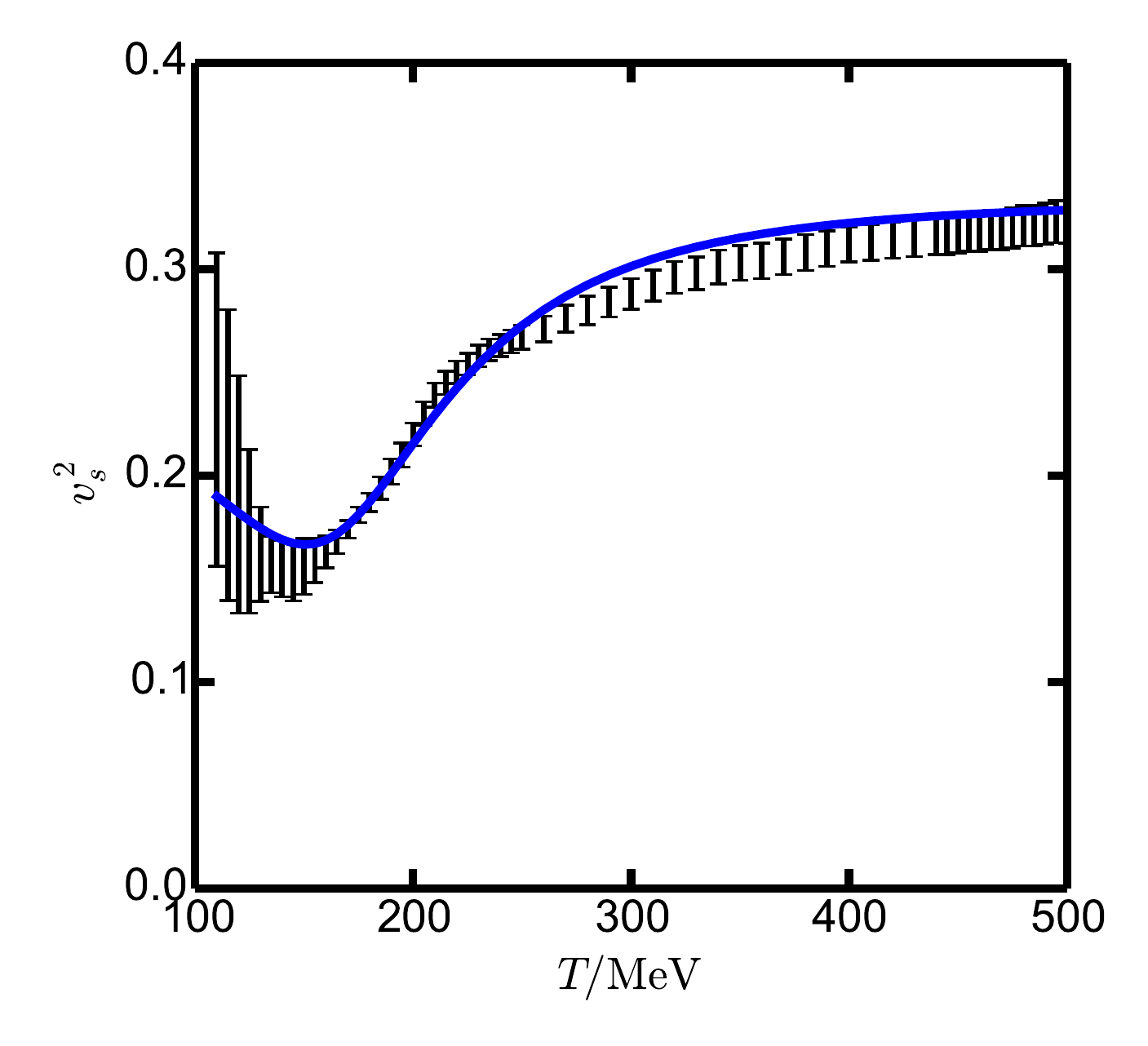} 
\includegraphics{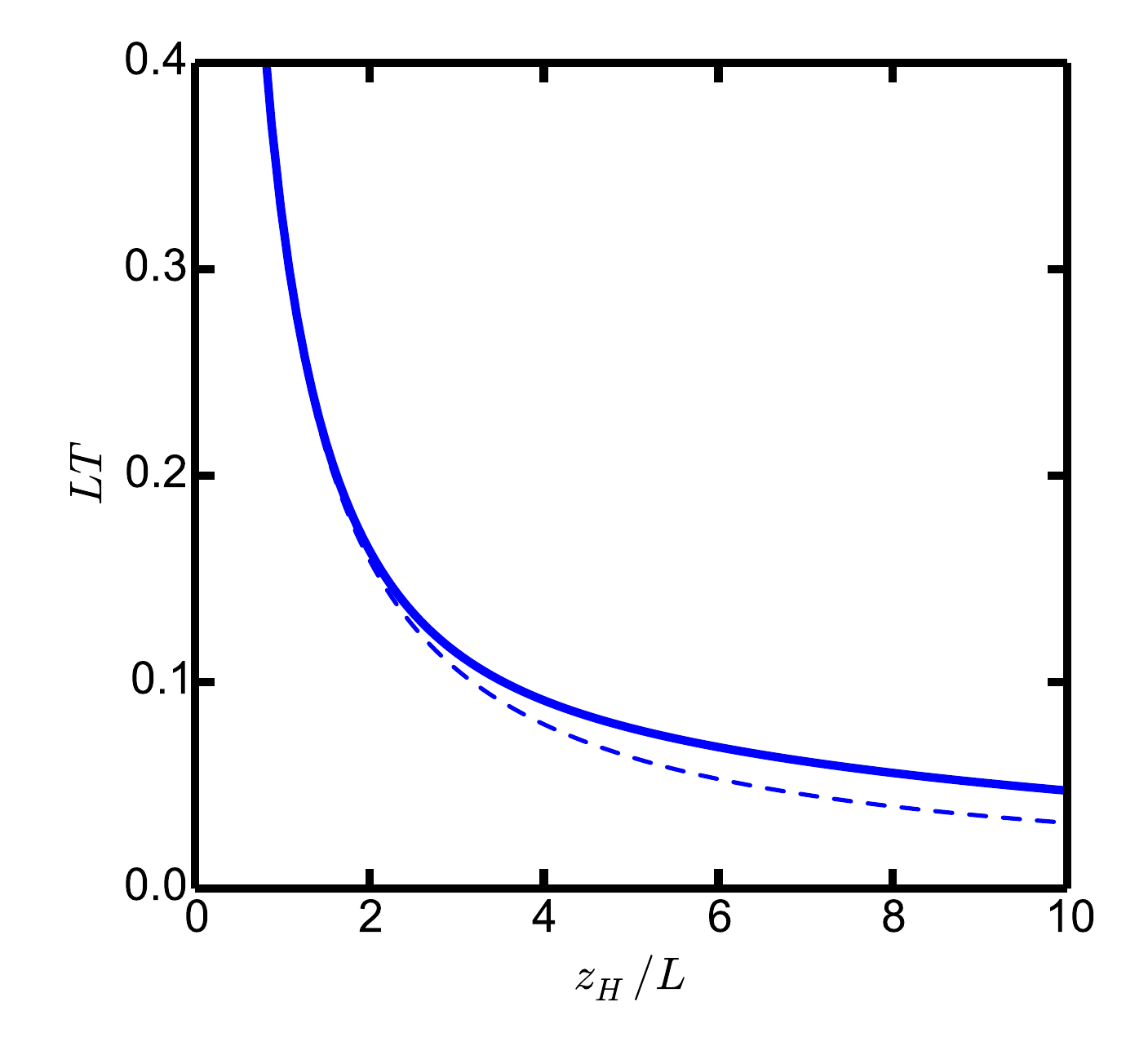}
}
\caption{Adjustment of the dilaton potential (\ref{eq.A2}) to lattice QCD data
\cite{Borsanyi:2013bia,Bazavov:2014pvz}, in particular, to the velocity of sound squared,
$v_s^2 = \frac{\dd \log T}{\dd \log s}$
as a function of temperature (left panel). Note that the dip in sound velocity
is characteristic for the cross-over. 
The resulting temperature $L T(z_H)$ as a function of $z_H / L$ 
is exhibited in the right panel by the solid curve; 
the dashed curve is for $T(z_H) = 1 / \pi z_H$. 
\label{fig:3}
}
 \end{figure}

Deforming the AdS metric by putting a Black Hole with horizon at $z_H$
yields the metric for the infinitesimal line elements squared 
\begin{equation} \label{eq:3}
\dd s^2 = \exp\{ A(z, z_H)\} 
\left[f(z, z_H) \, \dd t^2 - \dd \vec x^2 - \frac{\dd z^2}{f(z, z_H)} \right] ,
\end{equation}
where $f(z, z_H) \vert_{z = z_H} = 0$ is a simple zero. Identifying the
Hawking temperature $T(z_H) = - \partial_z f(z, z_H) \vert_{z = z_H} /4 \pi$
with the temperature of the system at bulk boundary $z \to 0$, 
and the attributed Bekenstein-Hawking entropy density 
$s(z_H) = \frac{2 \pi}{\kappa} 
\times
\exp\{ \frac32 A(z, z_H)\vert_{z=z_H} \}$
one describes holographically the thermodynamics.
$f = 1$ at $T = 0$ is attributed to vacuum.
The gravity-dilaton background is determined by the action in the Einstein frame 
\begin{equation}
S = \frac{1}{2 \kappa} \int \dd^4 x \, \dd z \sqrt{g_5}
\left[R - \frac12 (\partial_z \phi)^2 - V(\phi) \right] ,
\end{equation} 
where $R$ stand for the curvature invariant and $\kappa = 8 \pi G_5$. 
(For our purposes, the numerical values of $\kappa$ and $G_5$ as well as $k_V$
in (\ref{eq:1}) are irrelevant.)
The dilaton potential $V(\phi)$ is the central quantity \cite{Zollner:2018uep}.
We use a simple three-parameter ansatz
\begin{equation} \label{eq.A2}
- L^2 V = 12 \cosh(\gamma \phi) + \phi_2 \phi^2 + \phi_4 \phi^4
\end{equation}
to  find from the field equations and equation of motion for the metric (\ref{eq:3})
\begin{eqnarray}
A'' &=& \frac12 A'^2 - \frac13 \phi'^2 , \label{A:3}\\
f'' &=& - \frac23 A' f' ,  \label{A:4}\\
\phi'' &=& -\left( \frac23 A' +\frac{f'}{f} \right) \phi' + 
\frac{1}{f} e^A \partial_\phi V \label{A:5}
\end{eqnarray}
the suitable coefficients $(\gamma, \phi_2, \phi_4) = (0.568, -1.92, -0.04)$
together with $L^{-1} = 1.99$~GeV
which deliver a satisfactory description of the lattice QCD$_{2+1}$(phys) data
\cite{Borsanyi:2013bia,Bazavov:2014pvz}, see Fig.~\ref{fig:3}.
The prime means differentiation w.r.t.\ $z$
in (\ref{A:3}) - (\ref{A:5}), and boundary conditions are 
$A(z \to 0) \to -2 \log (z/L)$, 
$\phi(0) = 0$, $\phi'(0) = 0$,
$f(0) = 1$, $f(z_H) =0$. \\
Despite the conformal dimension $\Delta$ (from $\Delta (\Delta - 4)= L^2 m_\phi^2$ leading to $\Delta= 2+\sqrt{4-2\phi_2-12\gamma^2}= 3.9$) of the dual operator to $\phi$ is not four, 
we denote here the bulk scalar field $\phi$ as ``dilaton'', thus following the nomenclature, e.g.\ in \cite{Finazzo:2014cna,Finazzo:2013efa,Rougemont:2015wca}.

\end{document}